\documentclass{aa}  
\usepackage{graphicx}
\usepackage[varg]{txfonts}
\usepackage{subfig}
\usepackage[labelfont=bf]{caption}

\DeclareMathOperator{\sech}{sech}
\newcommand{\kms}{\,km\,s$^{-1}$}

\usepackage{hyperref}
\hypersetup{colorlinks=true,linkcolor=red,citecolor=blue,urlcolor=magenta}

\begin{document}

   \title{The volumetric star formation law in the Milky Way}

%    \subtitle{}
   \authorrunning{C. Bacchini et al.}
   \author{C. Bacchini\inst{1,2,3},
          F. Fraternali\inst{2,1},
          G. Pezzulli\inst{4},
          A. Marasco\inst{2,5},
          G. Iorio\inst{6}, \and
          C. Nipoti\inst{1}          
          }

   \institute{Dipartimento di Fisica e Astronomia, Universit\`{a} di Bologna, via P. Gobetti 93/2, 40129 Bologna, Italy\\
   \email{cecilia.bacchini@unibo.it}
   \and
    Kapteyn Astronomical Institute, University of Groningen, Postbus 800, 9700 AV Groningen, The Netherlands
   \and
   INAF - Osservatorio di Astrofisica e Scienza dello Spazio di Bologna, via Gobetti 93/3, I-40129 Bologna, Italy
   \and
   Department of Physics, ETH Zurich, Wolfgang-Pauli-Strasse 27, 8093 Zurich, Switzerland
   \and 
    ASTRON, Netherlands Institute for Radio Astronomy, Oude Hoogeveensedijk 4, 7991 PD, Dwingeloo, The Netherlands
   \and
   Institute of Astronomy, University of Cambridge, Madingley Road, Cambridge CB3 0HA, UK
}

   \date{}

% \abstract{}{}{}{}{} 
% 5 {} token are mandatory
 
  \abstract{
  Several open questions on galaxy formation and evolution have their roots in the lack of a universal star formation law, that could univocally link the 
  gas properties, e.g. its density, to the star formation rate (SFR) density. 
  In a recent paper, we used a sample of nearby disc galaxies to infer the volumetric star formation (VSF) law, a tight correlation between the gas and the SFR volume densities 
  derived under the assumption of hydrostatic equilibrium for the gas disc. 
  However, due to the dearth of information about the vertical distribution of the SFR in these galaxies, we could not find a unique slope for the VSF law, but two 
  alternative values. 
  In this paper, we use the scale height of the SFR density distribution in our Galaxy adopting classical Cepheids (age~$\lesssim 200$ Myr) as tracers of star formation. 
  We show that this latter is fully compatible with the flaring scale height expected from gas in hydrostatic equilibrium. 
  These scale heights allowed us to convert the observed surface densities of gas and SFR into the corresponding volume densities. 
  Our results indicate that the VSF law $\rho_\mathrm{SFR} \propto \rho_\mathrm{gas}^\alpha$ with $\alpha \approx 2$ is valid in the Milky Way as well as in nearby disc galaxies.}

   \maketitle
%
%-------------------------------------------------------------------
\section{Introduction}\label{sec:intro}
Sixty years ago, \cite{1959Schmidt} theorised the first star formation law for the Milky Way (MW): a power-law linking the star formation rate (SFR) and the atomic gas volume 
densities $\rho_\mathrm{SFR} \propto \rho_\mathrm{HI}^n$. 
He estimated $2<n<3$ from the HI and young stars distributions in our Galaxy.
To date, many efforts have gone into finding a universal relation between gas and SFR densities among all types of star-forming galaxies. 
We could divide the star formation laws proposed in the literature into three main groups, according to the physical quantities and scales considered. 

The so-called \emph{global} Schmidt-Kennicutt (SK) law involves the surface densities of gas (HI+H$_2$; $\Sigma_\mathrm{gas}$) and SFR ($\Sigma_\mathrm{SFR}$) averaged over the 
whole star-forming disc. 
This correlation was proposed by \cite{1998Kennicutt} using a sample of regularly star-forming disc galaxies and starbursts, and it reads 
$\Sigma_\mathrm{SFR} \propto \Sigma_\mathrm{gas}^N$ with $N \approx 1.4$.  
Recently, \cite{2012KennicuttEvans} showed that the MW is compatible with the integrated SK law, while low surface brightness galaxies depart from the relation 
\citep[see also][]{2019delosReyesKennicutt}. 

For spatially resolved galaxies, it is possible to derive the \emph{resolved} SK law \citep[e.g.][]{1989Kennicutt,2001MartinKennicutt}, which involves the gas and the SFR surface 
densities measured either in kpc or sub-kpc regions, or their  radial profiles (i.e. azimuthal averages). 
However, like its integrated version, this correlation seems to break in low density environments, as found by several authors in dwarf galaxies and the outskirts of spirals 
\citep[e.g.][]{,2007Kennicutt,2011Bolatto,2014Dessauges}. 
This is often ascribed to a drop in the star formation efficiency at a threshold density of about 10 M$_\odot$ pc$^{-2}$ \citep[e.g.][]{2004Schaye,2008Leroy,2008Bigiel,2010Bigiel}; 
however the physical explanation for this behaviour is still a matter of debate (e.g. \citealt{2014Krumholz_rev} and references therein).  
In our Galaxy, it is unclear whether the index of the resolved SK law is 1.4 \citep[e.g.][]{2012FraternaliTomassetti} or higher \citep{2002Wong,2003Boissier,2006Misiriotis}, 
which may be an indication of the presence of the break \citep{2017Sofue}. 

Given that stars are thought to form from cold and dense gas, the SFR surface density is expected to correlate with the molecular gas surface density, following some 
\emph{molecular} star formation law. 
This correlation is observed in high gas density regions of spiral galaxies, altought its index has not been firmly established yet. 
In fact, some authors found a linear correlation \citep[e.g.][]{2008Bigiel,2011Schruba,2012Marasco}, while others derived an index around 1.4 
\citep[e.g.][]{2002Wong,2004Heyer,2007Kennicutt,2011Liu}. 
In our Galaxy, \cite{2006Luna} investigated the molecular SK law using the SFR density traced by the far infrared emission (i.e. dust heated by massive young stars), finding a 
power-law with index of $1.2 \pm 0.2$. 
However, \cite{2012KennicuttEvans} showed that the H$_2$ surface density drops faster with radius than the SFR distribution derived using HII regions 
\citep{2006Misiriotis,2017Sofue}. 
Moving to much smaller spatial scales, \cite{2010Lada} found a linear correlation between the number of young stellar objects in Galactic molecular clouds and the mass of 
dense gas above an extinction threshold of 0.8 magnitudes in $K$ band, corresponding to about 116 M$_\odot$ pc$^{-2}$. 
The origin of this correlation is unclear, as it could be a consequence of the scaling relation between mass and size of molecular clouds \citep{2013Lada}. 

In our previous paper \citep[][hereafter B19]{2019Bacchini}, we proposed a new volumetric star formation (VSF) law, a tight correlation between the SFR and the gas (HI+H$_2$) 
volume densities derived for a sample of 12 nearby star-forming galaxies. 
The conversion of the observed surface densities into volume densities requires the knowledge of the gas and SFR scale heights. 
In particular, the scale heights of the HI and H$_2$ components were computed assuming the vertical hydrostatic equilibrium in the galactic potential. 
A key feature of this approach is that it takes into account radial variations of the gas scale height (also called \emph{flaring}) and the consequent non-linear conversion 
between the observed surface density and the intrinsic volume density \citep[see also][]{2015Elmegreen,2018Elmegreen}. 
In the absence of observational measurements of the radial variation of the SFR vertical distribution, we decided to make two extreme assumptions for the SFR scale height. 
The first consisted in assuming a constant value for the entire disc (and the same for all galaxies), while the second was based on the idea that the SFR scale height is proportional 
to that of the most abundant gas phase, whether atomic or molecular. 
Clearly, these definitions for the SFR scale height led to two different radial profiles for the SFR volume density. 
In both cases we found a tight power-law relation, but with different indexes, $1.34 \pm 0.03$ and $1.91 \pm 0.03$.

In this work, we show that the issue about the SFR scale height can be overcome in the MW, where the 3D structure of the tracers of recent star formation can be 
directly retrieved from observations, and we assess the validity of the VSF law in our Galaxy. 
Sect.~\ref{sec:model} describes the model of the gas distribution and defines the volume densities. 
Sect.~\ref{sec:data} presents the measurements of the distributions of gas and SFR that we took from the literature. 
Then, our results are presented in Sect.~\ref{sec:results} and discussed in Sect.~\ref{sec:discussion}. 
Finally, Sect.~\ref{sec:sumconc} summarises the work and draws the main conclusions. 

\section{Volume densities}\label{sec:model}
In this paper, we apply to the MW the same approach proposed in B19 for external galaxies, with the exception of the SFR scale height. 
For the sake of clarity, we briefly summarise the adopted methods in the following.

\subsection{Distribution of gas in hydrostatic equilibrium}\label{sec:model_he}
We assume that the HI and H$_2$ discs are in vertical hydrostatic equilibrium in the total gravitational potential, which consists of a dark matter halo, 
a stellar bulge, a thin and a thick stellar disc, plus the contribution of the gas self-gravity. 
In general, for a given gravitational potential and gas density profile, it is possible to calculate the scale height of the gaseous component once its velocity 
dispersion ($\sigma$) is known, assuming that the pressure is $P = \rho \sigma^2$.  
Indeed, the density distribution can be written as
\begin{equation}\label{eq:rho_hydro}
 \rho_i(R,z) = \rho_i(R,0) \exp \left[ - \frac{\Phi(R,z)-\Phi(R,0)}{\sigma_i^2} \right] \, , 
\end{equation}
where $i$ stands for HI or H$_2$, $\rho_i(R,0)$ is the volume density in the midplane, and $\Phi$ is the total gravitational potential. 

We calculate the scale height via numerical integration using the software \textsc{Galpynamics}\footnote{\url{https://github.com/iogiul/galpynamics}} \citep{2018Iorio}, 
that uses an iterative algorithm to account for the gas self-gravity. 
In pratice, the code first calculates the external potential plus the contribution of a razor-thin gas distribution from a given parametric mass model. 
A first guess of the scale height is estimated by fitting a Gaussian profile \citep[e.g.][]{1995Olling,2009Koyama} to the gas distribution resulting from Eq.~\ref{eq:rho_hydro}. 
The \emph{scale height} ($h$) is defined as the standard deviation of such profile. 
Then, $\Phi$ is updated with the potential of the new gas distribution, which includes the scale height found in the previous step, and a second estimate of the scale height is 
obtained by the Gaussian fitting. 
This procedure is iterated until two successive calculations differ by less than a given tolerance factor. 

Therefore, the necessary ingredients to calculate the scale heights of HI and H$_2$ are their surface densities and velocity dispersions (see Sect.~\ref{sec:gas_dist_kin}), and a 
parametric mass model of the Galaxy components. 
In particular, we adopted the models for the stellar and dark matter distributions by \cite{2011McMillan,2017McMillan}, which take into account observational requirements on the 
kinematics of gas, stars, and masers, and on the total mass of the Galaxy out to 50 kpc.

\subsection{Definitions of volume densities}\label{sec:model_defvold}
We define the gas volume density in the Galaxy midplane as
\begin{equation}\label{eq:rho_gas}
\begin{split}
 \rho_\mathrm{gas}(R,0) & = \rho_\mathrm{HI}(R,0) + \rho_\mathrm{H_2}(R,0) \\
  & = \frac{\Sigma_\mathrm{HI}(R)}{\sqrt{2 \pi} h_\mathrm{HI}(R)} + \frac{\Sigma_\mathrm{H_2}(R)}{\sqrt{2 \pi} h_\mathrm{H_2}(R)} \, ,
\end{split}
\end{equation}
where $\rho_\mathrm{HI}(R,0)$ and $\rho_\mathrm{H_2}(R,0)$ are the volume densities of HI and H$_2$ in the midplane, and $\Sigma_\mathrm{HI}(R)$ and $\Sigma_\mathrm{H_2}(R)$ 
are the corresponding radial profiles of the surface densities. 
The last equality in Eq.~\ref{eq:rho_gas} holds under the assumption of a Gaussian vertical profile for HI and H$_2$, and $h_\mathrm{HI}$ and $h_\mathrm{H_2}$ are the standard 
deviations of such profiles. 
These latter were calculated using the procedure described in Sect.~\ref{sec:model_he} based on the assumption of hydrostatic equilibrium. 

We assumed that the SFR is distributed in a disc with surface density $\Sigma_\mathrm{SFR}(R)$ and scale height $h_\mathrm{SFR}(R)$. 
Hence, the volume density of SFR in the miplane is 
\begin{equation}\label{eq:rho_sfr}
\rho_\mathrm{SFR}(R,0)=\frac{\Sigma_\mathrm{SFR}(R)}{\sqrt{2 \pi} h_\mathrm{SFR}(R)} \, .
\end{equation}
The main difference with respect to B19 is that, in this work, we measured $h_\mathrm{SFR}(R)$ from observations. 

\section{Data}\label{sec:data}
\subsection{Gas distribution and kinematics in the Milky Way}\label{sec:gas_dist_kin}
Several works in the literature studied the gas distribution, i.e. its surface density, volume density, and scale height, in our Galaxy adopting the 
kinematic distance method \citep{1957Westerhout}. 
This latter relies on an assumed model of the Galactic rotation curve to transform the line-of-sight velocity into a distance from the Solar position. 
The derived distances can then be used to obtain a full 3D reconstruction of the gas component. 
This method has been successfully and widely employed to map the gas densities in the entire Galaxy, but it is affected by the so-called \emph{near-far problem} within the 
Solar circle \citep[e.g.][]{1974Burton,2017Marasco}: the same line-of-sight velocity can be associated with two opposite distances, one between the observer and the tangent point,
defined as the location where the line-of-sight is perpendicular to R, and one beyond it. 

In this work, we decided to derive the volume densities from the surface densities in the literature using the scale height calculated with the hydrostatic equilibrium. 
This approach allows us to compare, in a consistent way, the VSF law in the MW with that obtained in B19. 
In Appendix~\ref{ap:gas_distrib}, we discuss the difference between the volume density and the scale heights estimated in the literature and those 
derived using the hydrostatic equilibrium.

\subsubsection{Inside the Solar circle}\label{sec:gas_dist_kin_inner}
\cite{2017Marasco} studied the distribution and kinematics of the gas inside the Solar circle through a novel approach that models the observed emission of HI and CO, 
overcoming the near-far problem. 
They assumed that the gas is in circular motion and divided the Galaxy in concentric and co-planar rings described by rotation velocity, velocity dispersion, scale height, 
and miplane volume density. 
Then, they used a Bayesian method to fit these four parameters to the HI and the CO line emission from the Leiden-Argentine-Bonn survey \citep{2005Kalberla} and 
the CO(J=1$\rightarrow$0) survey of \cite{2001Dame}. 
This model also takes into account the extra-planar gas contribution, which was included as an additional HI component with both radial and vertical infall motions, and a lagging 
rotational velocity with respect to the HI in the midplane\footnote{The extra-planar gas is a faint layer of HI, observed both in the MW and in nearby galaxies 
\citep{2001Fraternali,2007Oosterloo,2013Gentile,2019Marasco}, that is likely generated by the galactic fountain flow \citep{2017Fraternali}. 
This component rotates slower with respect to the midplane gas and reaches heights of a few kpc above the disc. 
If not taken into account, the extra-planar gas can lead to a slight overestimate of the scale height ($\sim$ 20\% for our Galaxy; see \citealt{2017Marasco}).} 
\cite[see][]{2011Marasco}. 
Their best-fit model can reproduce in detail the gas distribution and kinematics of the receding and approaching quadrants of the Galaxy. 
The assumption of pure circular orbits does not hold though in the innermost 3 kpc, where the bar gravitational potential makes the gas distribution non-axisymmetric and induces 
non-circular motions \citep{1991Binney,2015SormaniI,2015SormaniIII,2019Armillotta}. 
We excluded the region $R<3$ kpc in this work as also our model assumes axisymmetry. 
The profiles of the surface density, the volume density, and the scale height provided by \cite{2017Marasco} show three peaks, that could be related to the intersection of the 
line of sight with spiral arms, where the density is above the mean value. 
Therefore, we smoothed all the profiles, including those of the velocity dispersion, from a resolution of 0.2 kpc to 1.0 kpc (see Appendix~\ref{ap:Sigmagas_compare}). 

\subsubsection{Beyond the Solar circle} \label{sec:gas_dist_kin_outer} 
We derived an averaged profile for $\Sigma_\mathrm{HI}$ from the measurements by \cite{1998BinneyMerrifield}, \cite{2003NakanishiSofue}, and 
\cite{2006Levine}, who all used the kinematic distance method but assumed slightly different Galactic rotation curves\footnote{We did not include \cite{2008KalberlaDedes} 
measurements in the estimate of our fiducial $\Sigma_\mathrm{HI}$ as their density profile is more than a factor 2 higher than the other estimates in the literature 
(see Appendix~\ref{ap:Sigmagas_compare}).}. 
For example, \cite{2006Levine} aimed to study the warp of the HI disc, which is present beyond $R\sim12$ kpc, thus assumed that the Galaxy circular speed is constant at 
220 \kms for $R>R_\odot$. 
On the other hand, \cite{2003NakanishiSofue} adopted the rotation curve from \cite{1998Dehnen}, which slightly decreases beyond $R_\odot$. 
These profiles of $\Sigma_\mathrm{HI}$ are approximately in agreement with the profile by \cite{2017Marasco} in the Solar vicinity (see Appendix~\ref{ap:Sigmagas_compare}). 
Similarly, we used the profiles from \cite{1998BinneyMerrifield} and \cite{2006NakanishiSofue} to calculate $\Sigma_\mathrm{H_2}$. 

Concerning the HI and H$_2$ velocity dispersion, there are no available measurements of their profiles beyond $R_\odot$, at least to our knowledge. 
However, several authors \citep[e.g.][]{2002Fraternali,2008Boomsma,2009Tamburro,2019Bacchini} showed that, in nearby spiral galaxies, $\sigma_\mathrm{HI}$ decreases with radius 
until it reaches values of about 8 \kms~and then remains roughly constant, in agreement with the outermost measurements, i.e. at $R_\odot$, by \cite{2017Marasco}. 
These latter also estimated $\sigma_\mathrm{HI}/\sigma_\mathrm{H_2} \approx 0.5$ within $R_\odot$, thus we decided to assign $\sigma_\mathrm{HI}=8 \pm 2$ \kms~and 
$\sigma_\mathrm{H_2}=4 \pm 1$ \kms~to all radii beyond $R_\odot$, which are tipical values for the atomic and the molecular phases (see \citealt{2016KramerRandall} and references 
therein).

\subsection{SFR distribution in the MW}\label{sec:sfr_distribution}
The SFR of our Galaxy can be estimated using different tracers of recent star formation. 	
\cite{2011ChomiukPovich} found that different measurements of the global SFR are consistent with $1.9 \pm 0.4$ M$_\odot$yr$^{-1}$, if rescaled to the Kroupa initial mass function 
 \citep{2003Kroupa} and stellar population models. 

\subsubsection{SFR surface density}\label{sec:sfr_surfacedensity}
We took as reference the work by \cite{2015Green}, who carefully collected a sample of 69 bright supernova remnants (SNRs)\footnote{In the MW, SNRs can be considered good 
tracers of recent ($\lesssim 50$ Myr) star formation events as, in Sbc galaxies, the rate of SNe Ia is $\sim 4-5$ times lower than the rate of SNe II and Ibc \citep{2011Li}.}
in order to avoid strong selection effects. 
He derived the radial distribution of SNRs and found that it is more concetrated towards the Galactic center with respect to the previous estimate by \cite{1998CaseBhattacharya}. 
We derived $\Sigma_\mathrm{SFR}(R)$ by normalising the radial profile of SNR surface density to the total SFR of the MW. 
In Appendix~\ref{ap:sfr_tracers_Sigma}, we show that the resulting $\Sigma_\mathrm{SFR}(R)$ is compatible with other estimates obtained with different tracers and methods, 
albeit with a large scatter. 

\subsubsection{SFR scale height}\label{sec:sfr_scaleheight}
To reliably measure $h_\mathrm{SFR}(R)$, we must select a tracer of recent star formation with accurate distance determination and a sample that is statistically significant. 
Therefore, we chose classical Cepheids (CCs), which are variable stars tipically younger than $\sim$200 Myr \citep[e.g.][]{2005Bono,2019Dekani} whose distance can be accurately 
determined thanks to the period-luminosity relation \citep[e.g.][]{1912LeavittPickering,2000Caputo,2018Ripepi}. 
We note that SNRs and CCs are not perfectly coeval (age gap 50-100 Myr), but the dynamical processes that could modify the distribution of a population of stars with 
respect to the parent gas (e.g. radial migration) are effective on time-scales much longer than this age gap \citep{2014Sellwood}. 
Moreover, the stellar discs of star-forming galaxies grow in radius (inside-out growth) of $\sim 3$\% in 1 Gyr \citep{2001MunozMateos,2015Pezzulli}.
Hence, we expect that both SNRs and CCs well represent the parent gas distribution (see also Fig.~\ref{fig:Sigma_h_SFR} for a comparison between tracers of 
different ages).

In a recent paper, \cite{2019Chen} collected data for 1339 CCs with distance accurancy of 3-5\% from the Wide-field Infrared Survey Explorer  
catalogue of periodic variables and from the \emph{Gaia} Data Release 2 in order to study the warp of the Galactic disc. 
They found that the warp seen in the distribution of CCs is compatible with that determined using pulsars \citep{2004Yusinov} and atomic gas \citep{2006Levine}. 
After subtracting the warp contribution, these authors found evidence of a flare in the $z$-distribution of CCs compatible with that traced by red giant stars 
\citep{2018Wang} and HI \citep{1990Wouterloot} in the MW (see Appendix~\ref{ap:hgas_compare}).

We studied the radial profile of the scale height using the residual $z$-coordinates, i.e. with the warp removed, of the CCs provided by \cite{2019Chen} 
(see Appendix~\ref{ap:sfr_tracers_h} for details). 
Fig.~\ref{fig:h} shows the scale height of CCs: we clearly see that there is a flaring. 
In particular, the scale height is 100 pc at the Solar position and increases with radius, reaching about 500 pc at $R \approx$18 kpc. 
This is a strong indication that the flaring SFR distribution is more realistic than that with a constant thickness, allowing us to disentangle between the two approaches adopted 
in B19 and choose the flaring $h_\mathrm{SFR}(R)$ instead of the constant one. 
In Appendix~\ref{ap:sfr_tracers_h}, the scale height of CCs is compared to the scale height of other SFR tracers provided in the literature, 
which are in agreement within the uncertainties. 

\section{Results}\label{sec:results}
\subsection{Scale heights of classical Cepheids and gas in hydrostatic equilibrium}\label{sec:h_gas_hydro}
Our first aim is to test the assumption made in B19, where we conjectured that the SFR scale height could be approximated by the weighted average of $h_\mathrm{HI}$ 
and $h_\mathrm{H_2}$ calculated with the hydrostatic equilibrium:
\begin{equation}\label{eq:h_SFR}
 h_\mathrm{gas}(R) = h_\mathrm{HI}(R) f_\mathrm{HI}(R) + h_\mathrm{H_2}(R) f_\mathrm{H_2}(R) \, ,
\end{equation}
where $f_\mathrm{HI}$ and $f_\mathrm{H_2}$ are the HI and the H$_2$ fractions with respect to the total gas. 

We derived the vertical distribution of HI and H$_2$ (Eq.~\ref{eq:rho_hydro}) as explained in Sect.~\ref{sec:model_he}. 
For consistency, we adopted the mass distribution for the gaseous components described in the Sect.~\ref{sec:gas_dist_kin} rather than those in \cite{2017McMillan} model. 
This choice has a negligible effect on the gravitational potential, as the gas is dynamically sub-dominant with respect to the stars and the dark matter. 
In Sect.~\ref{ap:hgas_compare}, we compare the scale heights of HI and H$_2$ obtained with the hydrostatic equilibrium with other determinations from previous studies. 

\begin{figure}
\includegraphics[width=1.\columnwidth]{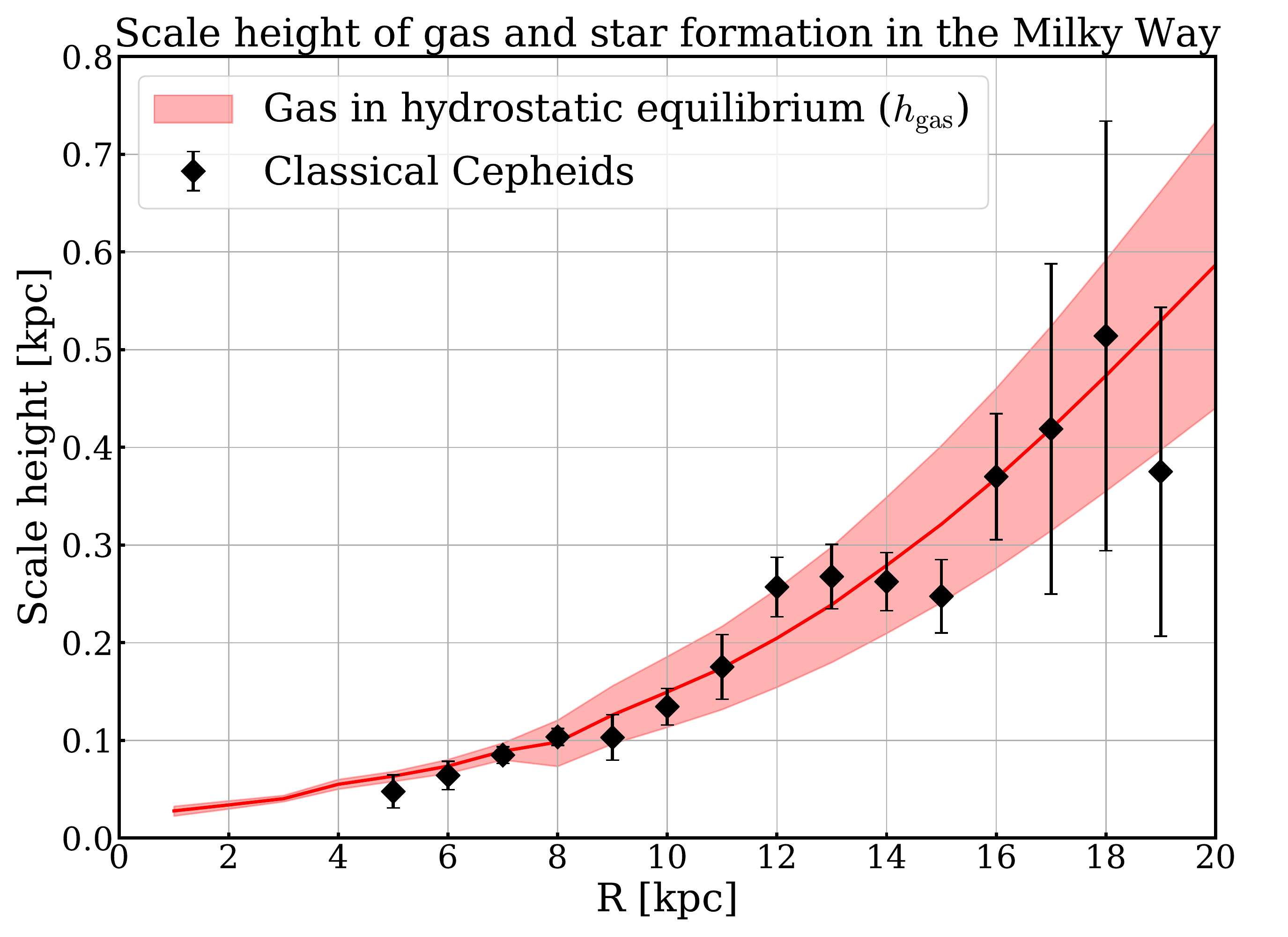}
\caption{Scale height of CCs (data from \citealt{2019Chen}) as a function of the Galactocentric radius (black diamonds). 
The red curve is the gas scale height (Eq.~\ref{eq:h_SFR}), i.e. the weighted average of the HI and the H$_2$ scale heights.}
\label{fig:h}
\end{figure}

In Fig.~\ref{fig:h}, the red curve shows the scale height of the gas defined by Eq.~\ref{eq:h_SFR} and the red band is the associated uncertainty calculated with Eq. E.7 in B19. 
Scale height profiles for the CCs and for the gas are in excellent agreement with each other, suggesting that the definition of $h_\mathrm{gas}(R)$ adopted in B19 
is optimal in describing the flaring of the SFR vertical distribution. 

\subsection{VSF laws in the MW}
Given the promising result discussed above, we investigated the location of the MW points on the volumetric correlations found in B19. 

\subsubsection{Total gas}
\begin{figure*}
\includegraphics[width=2.\columnwidth]{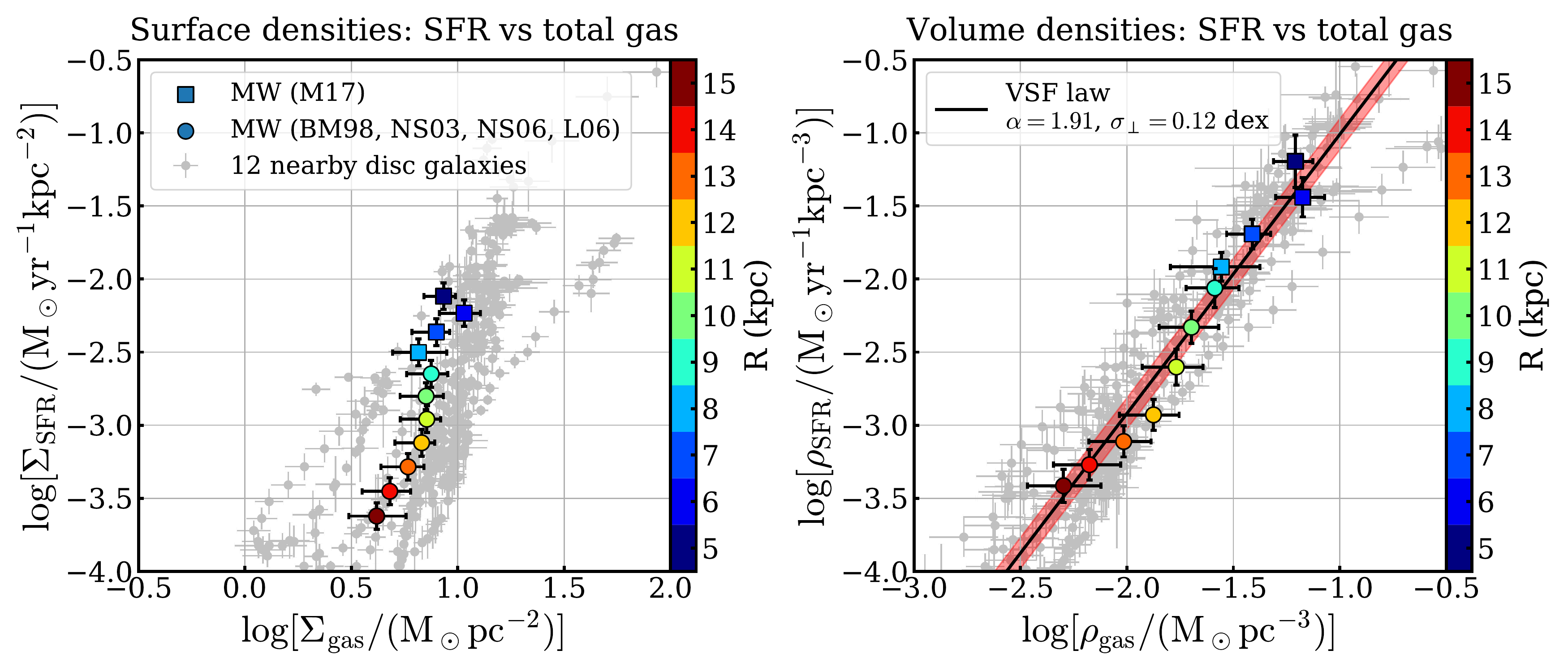}
\caption{Correlations between the surface density (left) and the volume density (right) of the gas and the SFR in the MW, colour-coded according to the Galactocentric 
radius. 
The squares indicate measurements for $R \leq R_\odot$ from \cite{2017Marasco}, while the circles are for $R>R_\odot$ from \cite{1998BinneyMerrifield}, 
\cite{2003NakanishiSofue,2006NakanishiSofue}, and \cite{2006Levine} (see text). 
The grey points are the corresponding quantites for the sample of 12 nearby disc galaxies (see Fig. 5 and Fig. 6 in B19 for the whole range of densities). 
The solid line in the right panel is the VSF law from B19 with its intrinsic scatter (red band).}
\label{fig:vsf}
\end{figure*}

In Fig.~\ref{fig:vsf}, we show the relations between $\Sigma_\mathrm{gas}$ and $\Sigma_\mathrm{SFR}$ (left panel) and $\rho_\mathrm{gas}$ and $\rho_\mathrm{SFR}$ (right panel), 
with the points color-coded according to the distance from the Galactic Center. 
We also include the sample of disc galaxies from B19 (grey points) in order to show that the MW follows the same trend as external galaxies. 
We note that the scatter is large in the surface-based panel, in particular for $\Sigma_\mathrm{gas}<$ 10 M$_\odot$ pc$^{-2}$. 
As shown in B19 (see left panel of Fig. 5b), a correlation close to $\Sigma_\mathrm{SFR} \propto \Sigma_\mathrm{gas}^{1.4}$ is visible at high densities, but some galaxies, 
including the MW, seem to follow a steeper relation with respect to the others. 

On the other hand, a different picture emerges from the right panel, where we can see that the MW volume densities follow remarkably well the VSF law with slope $\alpha=1.91$ 
(black solid line) and intrinsic scatter $\sigma=0.12$ dex (red band) found in B19 for nearby galaxies. 
We recall that $\rho_\mathrm{gas}$ for the MW (Eq.~\ref{eq:rho_gas}) was calculated using the scale heights derived with the hydrostatic equilibrium, consistently with the 
analysis done in B19 for nearby galaxies. 
Instead, $\rho_\mathrm{SFR}$ for the MW was estimated through Eq.~\ref{eq:rho_sfr} adopting the scale height of CCs, 
and not with Eq.~\ref{eq:h_SFR} as done for external galaxies. 
We discuss the relevance of this result in Sect.~\ref{sec:discussion}. 
In Appendix~\ref{ap:vsflaw_others}, we compare, for completeness, the VSF law in Fig.~\ref{fig:vsf} with the volume density of gas and SFR estimated using other measurements 
in the literature. 

\subsubsection{Atomic gas}
In B19, we found that there is a surprisingly tight correlation with slope between 2.1 and 2.8 involving the atomic gas and the SFR volume densities. 
This is different with respect to the results obtained by other authors using the corresponding surface densities, that seems to be completely uncorrelated 
\citep[e.g.][]{2010Bigiel,2011Schruba}. 

\begin{figure*}
\includegraphics[width=2.\columnwidth]{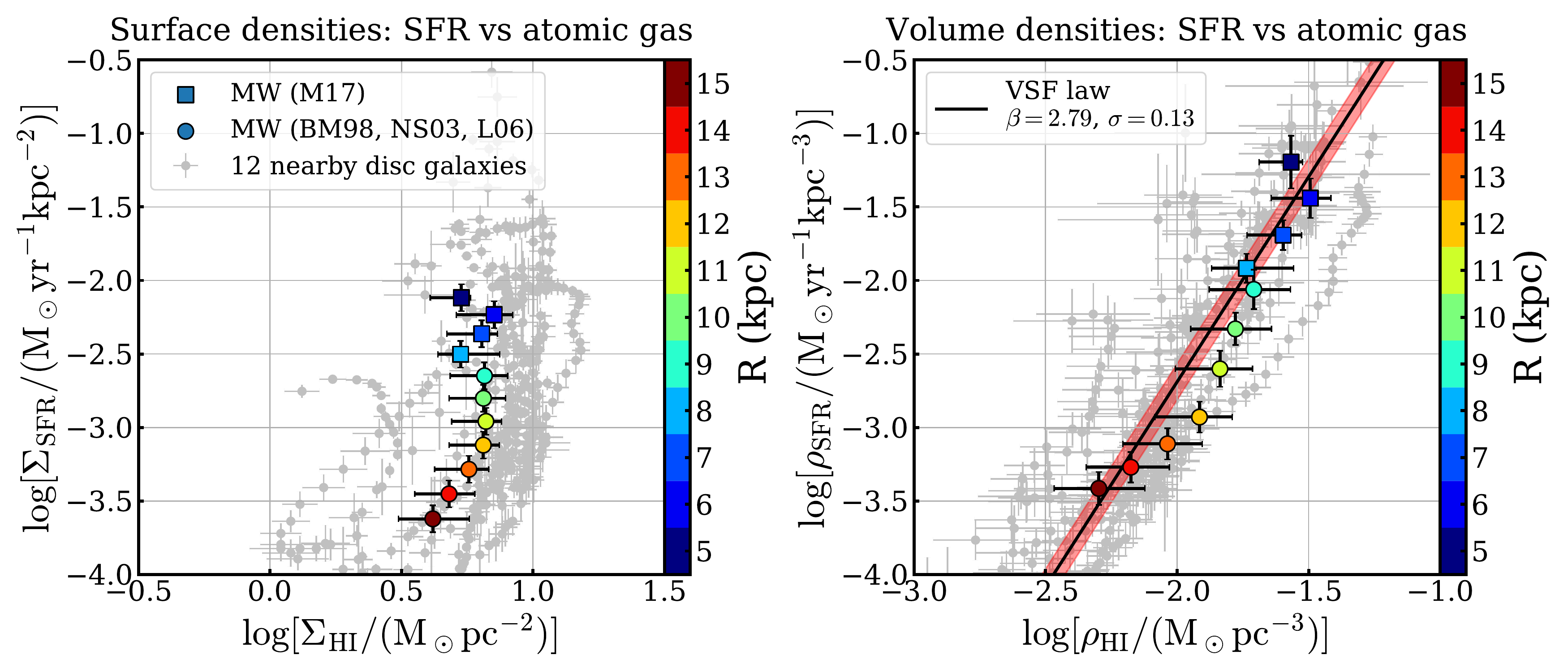}
\caption{Same as Fig.~\ref{fig:vsf} but in this case the gas densities are of the atomic gas only.}
\label{fig:vsf_hi}
\end{figure*}

The left panel of Fig.~\ref{fig:vsf_hi} shows the surface densities in the MW (points colour-coded using $R$) and in the sample of nearby disc galaxies of B19 (grey points). 
The MW is consistent with the other galaxies also in the plane $\Sigma_\mathrm{HI}$--$\Sigma_\mathrm{SFR}$, as in the case of total gas, and it is clear that there is very 
weak or no correlation between these two quantities. 

On the contrary, the right panel shows that the MW volume densities of atomic gas and SFR correlate, following remarkably well the VSF law with slope $\beta=2.79 \pm 0.08$ 
found for external galaxies. 
The validity of this relation in the MW confirms the link between atomic gas and star formation. 
We discuss this apparently controversial result in Sect.~\ref{sec:phys_interp_totalgas}.

\subsubsection{Molecular gas}
\begin{figure*}
\includegraphics[width=2.\columnwidth]{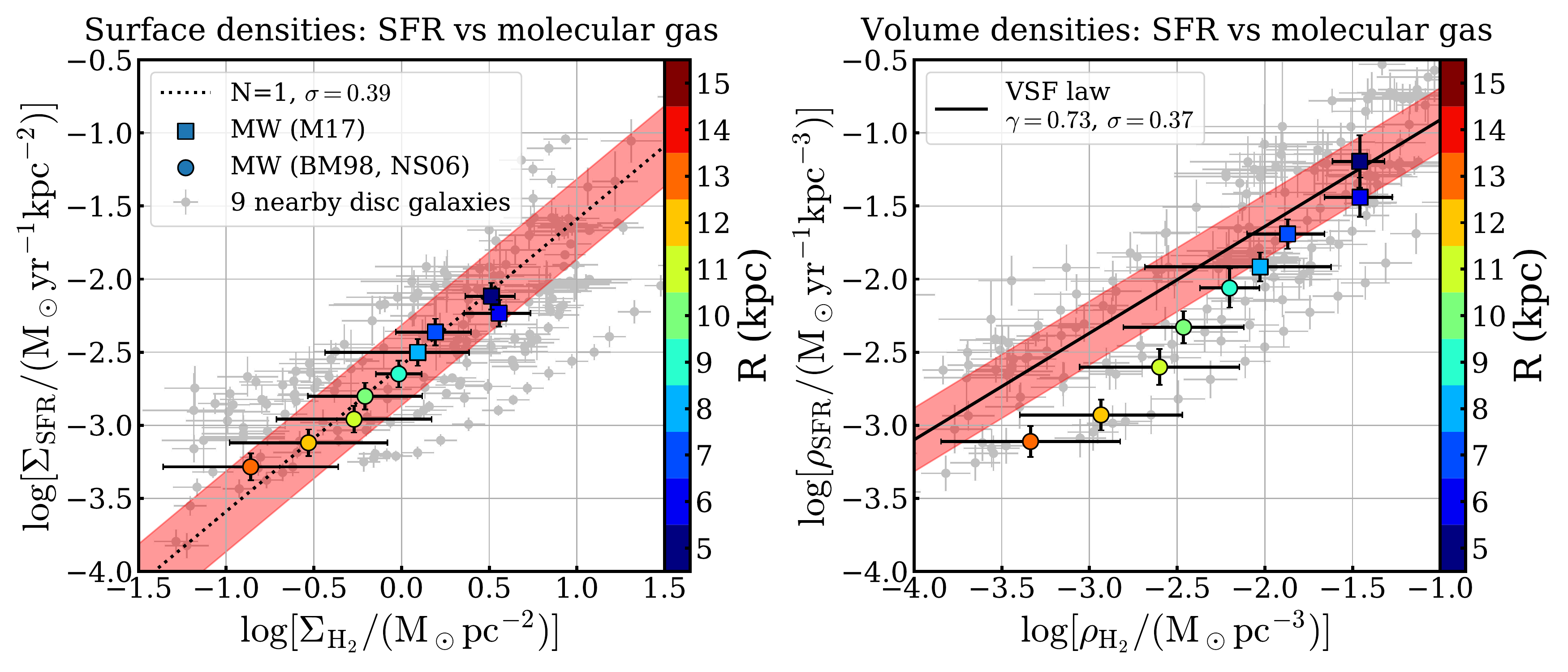}
\caption{Same as Fig.~\ref{fig:vsf} but in this case the gas densities are of the molecular gas only.}
\label{fig:vsf_h2}
\end{figure*}

The left panel of Fig.~\ref{fig:vsf_h2} shows the surface densities of molecular gas and SFR in the MW (coloured points) superimposed on those of the sample of disc galaxies 
in B19 (grey points). 
These latter were derived by \cite{2016Frank} using the CO conversion factor ($\alpha_\mathrm{CO}$) measured by \cite{2013Sandstrom}, which varies not only from galaxy to galaxy, 
but also with the galactocentric radius. 
The points of the MW are compatible with the relation obtained in B19 by fixing the slope to $\gamma=1$, as suggested by several authors 
\citep[e.g.][]{2002Wong,2007Kennicutt,2011Schruba,2012Marasco}. 
We note that the molecular gas is not detected beyond $R=13$ kpc, while $\Sigma_\mathrm{SFR}$ is still measured. 

Concerning the volume densities, the right panel of Fig.~\ref{fig:vsf_h2} shows that the MW points are compatible with external galaxies, given the large errorbars mainly due 
to the uncertainty on $\alpha_\mathrm{CO}$ \citep[see][]{2013Bolatto}. 
However, as we found in B19, the scatter of the relation is not significantly improved by the conversion from surface densities to volume densities. 
We also note that, by comparing Fig.~\ref{fig:vsf_h2} with Fig.~\ref{fig:vsf} and Fig.~\ref{fig:vsf_hi}, the scatter is larger in the case of molecular gas densities than 
in the case of atomic and total gas. 

\section{Discussion}\label{sec:discussion}
\subsection{Previous works on star formation laws in the Galaxy and nearby disc galaxies}
Several studies in the literature aimed to find a model of star formation that could reproduce the radial profile of the SFR in our Galaxy, given a gas distribution. 

For example, \cite{2003Boissier} adopted the Toomre criterion for the stability of galactic discs \citep{1964Toomre} proposed by \cite{1994Wang}, in order to account for 
the contribution of the stellar disc. 
They found that this criterion applies in the MW, but it has limited success in reproducing the profiles of the SFR surface densities in their sample of 16 disc galaxies. 
Moreover, the authors investigated the validity of the classical SK law: $\Sigma_\mathrm{SFR} \propto \Sigma_\mathrm{gas}^n$ \citep{1998Kennicutt} and of two modified versions, 
which make use of the galactic orbital time: $\Sigma_\mathrm{SFR} \propto \Sigma_\mathrm{gas}^n V_\mathrm{c}/R$ \citep{1975Ohnishi}, and of the stellar surface density: 
$\Sigma_\mathrm{SFR} \propto \Sigma_\mathrm{gas}^n (\Sigma_\mathrm{gas}+\Sigma_\star )^m $ \citep{1994DopitaRyder} 
\footnote{This formulation is almost equivalent to the so-called \emph{extended-Schmidt law} ($\Sigma_\mathrm{SFR} \propto \Sigma_\mathrm{gas} \Sigma_\star^{1/2}$), 
that was originally proposed by \cite{1975TalbotArnett} and then observationally studied by \cite{2011Shi,2018Shi} and \cite{2017Roychowdhury}.}, respectively. 
They found that both modified versions of the SK law work slightly better than the classical relation in both the MW and the external galaxies, but the scatter remains large. 
It is interesting to note that both modified SK relations are in some way included in our VSF law. 
Indeed, the orbital time and, in particular, the rotation curve of a galaxy depend on its gravitational potential, hence including this term in the correlation could 
partially account for the role of the gravitational pull in shaping the gas vertical distribution (see Eq.~\ref{eq:rho_hydro}). 
Moreover, the stellar mass component dominates the gravitational potential in the inner regions of disc galaxies, hence the scale height significantly depends on the 
stellar distribution. 

Our approach is based on the assumption that the vertical distribution of gas in galaxies is shaped by the hydrostatic equilibrium. 
This idea shares similarities with that proposed by \cite{2006BlitzRosolowsky}, hereafter BR06. 
These authors collected a sample of 14 nearby star-forming galaxies, including both dwarfs and spirals, and HI-rich and H$_2$-rich galaxies. 
They found that the ratio of the molecular to the atomic gas content correlates with the midplane pressure calculated using the hydrostatic equilibrium. 
Therefore, they proposed that this ``hydrostatic pressure'' regulates the SFR surface density, which has two formulations whether low-pressure (HI-dominated) environments or 
high-pressure (H$_2$-dominated) ones are considered. 
These results are somewhat different from ours as we do not find two regimes of star formation, but instead a monotonic relation that encompasses HI- and H$_2$-dominated regions.
We think that such differences may be, at least partially, explained by drawbacks in BR06's methodology. 
The most critical is that they neglected the dark matter component in the mass distribution calculation. 
As a consequence, their gravitational force and hydrostatic pressure are significantly underestimated, in particular in the outer regions of galaxies. 
Moreover, they ignored the radial gradient of the circular speed \citep[e.g.][]{1984Bahcall,1984BahcallCasertano,1995Olling}. 
This term is not negligible in the inner regions of galaxies, where most star formation takes place, and has an impact on the correct determination of the hydrostatic pressure. 
Finally, BR06 assumed that the gas velocity dispersion is constant with the galactocentric radius at 8 \kms, which is a factor 1.5-2 lower than the tipical 
values in the inner regions of galaxies \citep[e.g.][and B19]{2002Fraternali,2008Boomsma}. 

Recently, \cite{2017Sofue}, hereafter S17, used both the volume densities and the surface densities to study power-law correlations between the SFR and the atomic gas, 
the molecular gas, and the total gas in our Galaxy. 
He found that both the index and the normalisation of all the power-laws vary with the Galactocentric radius and that all the relations tend to steepen in the outer regions of 
the Galaxy ($8~\mathrm{kpc}<R\leq 20~\mathrm{kpc}$) with respect to the inner regions ($0~\mathrm{kpc}\leq R \leq 8~\mathrm{kpc}$), suggesting the existence of a density threshold
\footnote{The variation of the index and the normalisation was measured also on shorter scales by dividing the Galaxy in annulii of 2 kpc width.}. 
In agreement with our findings, he obtained that the HI-SFR volume density correlations are steeper than those involving the total gas ones, which are in turn steeper than the 
H$_2$-SFR relations. 
Despite evident similarities, there are significant differences between S17's and our approach, as we have already discussed in Sect.~\ref{sec:gas_dist_kin} and 
Appendix~\ref{ap:gas_distrib} concerning the gas distribution. 
In addition, the SFR distribution adopted by S17 was derived in a previous paper \citep{2017PASJSofue} using HII regions and adopting the kinematic distance 
method to infer their positions in the Galaxy. 
These authors decided to convert the surface density of SFR to the volume density using a constant value for $h_\mathrm{SFR}$, as they found that the scale height of HII 
regions is $\sim 50$ pc within $R\leq10$ kpc, showing clear flaring only beyond 15 kpc. 
This result is in contrast with that of \cite{2004Paladini}, who also studied the distribution of HII regions, and with the measurements of the scale height of other 
SFR tracers (see Appendix~\ref{ap:sfr_tracers_h}). 
Possibly, the $h_\mathrm{SFR}$ by \cite{2017PASJSofue} is contaminated by the uncertainties on the distance determination related the kinematic method, that we avoided 
using standard candels as CCs. 
Interestingly, S17 also measured the index of the power-laws considering the whole radial range, from 0 kpc to 20 kpc, and found $2.01 \pm 0.02$ for the correlation 
with the total gas volume density, $0.7 \pm 0.07$ for that with the molecular gas only, and $2.29 \pm 0.3$ for the atomic gas only, which are roughly consistent with our findings.

\subsection{Physical interpretation of the VSF law with total gas}\label{sec:phys_interp_totalgas}
In our VSF law, the SFR volume density is regulated by the square of the total gas volume density. 
In the following, we discuss possible interpretations of this correlation. 
The first issue to bear in mind is that the correlations that we found are valid on $\sim$kpc scales, but are likely not applicable to a single molecular cloud or filament. 
Moreover, the exact value of $\alpha$ remains slightly uncertain.  

Our VSF law can be written as 
\begin{equation}\label{eq:sfr_time}
 \rho_\mathrm{SFR} \propto \epsilon_\mathrm{sf} \frac{\rho_\mathrm{gas}}{\tau_\mathrm{sf}} \, ,
\end{equation}
where $\epsilon_\mathrm{sf}$ is a dimensionless efficiency parameter, usually assumed constant, and $\tau_\mathrm{sf}$ is some physically meaningful time-scale, which 
is different for different models and can have an explicit dependence on gas density.
A natural time-scale for star formation may be the free-fall time $\tau_\mathrm{ff} \propto \rho_\mathrm{gas}^{-1/2}$ \citep[e.g.][]{1977Madore,2012Krumholz}, which implies an 
index of 1.5. 
This is usually invoked to explain the SK relation with slope 1.4, with the implicit assumption of a constant scale height, both for the gas and the SFR distributions. 
In this work and in B19, we showed that these scale heights instead increase with radius, for the MW as well as nearby disc galaxies 
(see also \citealt{2008Abramova} and \citealt{2015Elmegreen,2018Elmegreen}). 
Moreover, our results indicate that the slope is closer to $\sim 2$ rather than 1.5, suggesting that another time-scale is driving star formation on kpc-scales or is involved 
in the process. 
Potentially interesting and physically relevant time-scales are the cooling time of warm gas ($T \approx 10^4$ K) and the time-scale to reach the equilibrium between the 
formation and distruction of H$_2$, as suggested by \citealt{2017Sofue} for example. 
We note that both these time-scales are inversely proportional to $\rho_\mathrm{gas}$ \citep[see for example][]{1979Hollenbach,2007CiottiOstriker,2014Krumholz_rev}, which, inserted in 
Eq.~\ref{eq:sfr_time}, would predict a VSF with index 2, in agreement with our findings.

Leaving aside these considerations about the slope of the VSF law, it is interesting to qualitatively interpret our results in the picture of a self-regulating star formation model. 
Our $\rho_\mathrm{gas}$ is the highest gas density (i.e. that at the midplane) at which the pressure-gravity balance holds (see Eq.~\ref{eq:rho_hydro} and Eq.~\ref{eq:rho_gas}). 
Therefore, it is probably a good estimate of the gas volume density (averaged on $\sim$kpc-scale) in star-forming clouds, as suggested by the small scatter of the VSF law. 
Turbulent motions are likely sustained by stellar feedback, whose strenght is expected to be proportional to the SFR density itself. 
The higher is the SFR density per unit volume, the more turbulent pressure helps thermal pressure in contrasting gravity. 
As a consequence, the gas $z$-distribution broadens and the scale height increases, thus the volume density in the midplane decrease and so does the SFR density. 
Then, the pressure support against gravity weakens because of the reduction of turbulence injection and the gas $z$-distribution narrows, increasing the gas volume 
densities in the midplane and, consequently, the SFR density. 
However, the influence of stellar feedback on the ISM turbulence is still a matter of debate \citep[e.g.][]{2009Tamburro,2019Utomo} and we leave a detailed analysis of this to 
future investigations. 

The above interpretation is similar to the self-regulating scenario proposed by \cite{2011Ostriker}, who assumed that star formation in spiral galaxies is regulated by the 
pressure support of gas turbulence against gravity (the radiation pressure may become dominant in the most extreme starburst regions). 
Investigating the compatibility of this model with our results also goes beyond the scope of this paper. 

The unexpected correlation between the SFR and the atomic gas volume densities suggests that the HI can be a good tracer of the star-forming gas over a broad range of densities, 
from dwarf to spiral galaxies. 
On the contrary, the scatter in the H$_2$-SFR relations is large, with no improvement from the conversion of surface densities to volume densities. 
In addition, the molecular gas, which is usually traced using CO emission, is often detected only in the inner regions of star-forming galaxies (including the MW). 
This may seems counter-intuitive, as star formation is observed to occur in molecular clouds. 
A possible explanation of the HI-SFR correlation is that molecular clouds form from the atomic gas and are rapidly swept away by stellar feedback, 
leaving only the parent atomic gas. 
Therefore, we could expect to see the HI-SFR correlation if the timescale for the formation of a new molecular clouds is longer than the lifetime of star formation 
tracers\footnote{The time-scale of molecular clouds formation depends on the physical mechanism that regulates the process. Probably, it is between a few $10^7$ yr 
and $10^8$ yr, subject to the ISM local conditions \citep[see for example][]{2007MccKeeOstriker}. Star formation tracers are characterised by time-scales that tipically range from 
a few $10^6$ yr to about $10^8$ yr \citep{2012KennicuttEvans}. Therefore, the comparison between the two times-scales is very uncertain.}.
In low-density and/or metal-poor regions, the CO emission likely becomes a bad tracer of the total molecular gas (H$_2$) \citep[e.g.][]{2012Schruba,2015Hunt,2017Seifried}. 
The existence of a HI-VSF law that extends to these environments seems then to indicate that HI can efficiently trace also the CO-dark H$_2$. 
Nevertheless, this correlation may also suggest that the atomic gas has an important, albeit non-trivial, role in star formation, and that the molecular gas is not always a 
prerequisite to create new stars \citep{2012GloverClark,2012KrumholzHI}.

\section{Summary \& conclusions}\label{sec:sumconc}
Star formation laws are undoubtedly of fundamental importance to understand galaxy formation and evolution. 
It is generally acknowledged that the formation of stars in galaxies is regulated by their gas reservoir, but studying the intrinsic distributions of the gas and the SFR 
is hampered by the difficulty of measuring the \emph{volume} densities in galaxies. 
The surface densities, observable in external galaxies, are affected by projection effects due to the flaring of the thickness of gas discs. 
Therefore, the relation between the surface density and the volumetric one is non-trivial and similar values of surface density can be measured where the volume density is low 
and the gas disc is thick, or viceversa. 

In B19, we used a sample of 12 nearby disc galaxies to derive the volume densities of the gas (HI and H$_2$) from the observed surface densities using the scale 
height calculated under the assumption of hydrostatic equilibrium in the galactic gravitational potential. 
We found that the gas and the SFR volume densities correlate, following a tight power-law relation, the VSF law, whose index is either $\sim$1.3 or $\sim$1.9 
depending on whether the scale height of the SFR distribution is assumed to be constant or flaring with the galactocentric radius. 

In this work, we investigated the VSF law in our Galaxy using the same method as for external galaxies, but with a crucial improvement. 
We used CCs as tracers of the recent star formation and thereby directly derived the thickness of their vertical distribution as a function of the Galactocentric 
radius. 
This allowed us to convert the SFR surface density to volume density and find a unique slope of the VSF law. 
Our main conclusions are the following.
\begin{enumerate}
 \item The vertical distribution of the SFR density flares with the Galactocentric radius and its scale height is fully compatible with the scale height of cold gas (HI+H$_2$) 
 calculated assuming the hydrostatic equilibrium. 
 \item We explored the correlations between the volume density of the SFR and the volume densities of the total gas, the atomic gas only, and the molecular gas only, 
 finding that the MW follows the same relations found in nearby disc galaxies. 
 \item The VSF law with total gas $\rho_\mathrm{SFR} \propto \rho_\mathrm{gas}^\alpha$ is the tightest correlation and $\alpha \approx 2$. 
\end{enumerate}

We stress that the flaring of gas thickness is significant and must be taken into account in the studies of gas distribution in galaxies, not only in dwarfs but also in 
spirals (see \citealt{2019Wilson} for an application to ultra-luminous infrared galaxies). 
The VSF law is described by a single index across the whole density range, which means that there is no volume density threshold and that the breaks previously observed 
in the resolved and integrated SK laws are due to the disc flaring, rather than to a drop of the star formation efficiency (see also \citealt{2015Elmegreen,2018Elmegreen}).

A physical interpretation of the VSF law is currently lacking and we hope that it will stimulate future investigations. 
Also the assumption of the hydrostatic equilibrium for the gas in galaxies should be tested, but measuring the vertical distribution of gas in galaxies is not an easy task. 

Our VSF laws are simple recipes for star formation that could be included in analytical or semi-analytical models of the formation and evolution of galaxies. 
Moreover, the VSF laws could be easily compared with the correlations between the SFR and the gas volume densities predicted by numerical simulations of 
star formation on small scales, with the advantage of avoiding the 2D projection to obtain the surface densities involved in the SK law.

\section*{Acknowledgements}
We are grateful to the anonymous referee for the useful comments that helped to improve the paper. 
We would like to thank X. Chen, T. Mackereth, and Y. Sofue for sharing their data and results. 
FF and AM thank Lotte Elzinga for her preliminary investigation of hydrostatic equilibrium in the Milky Way during her Bachelor project. 
GP acknowledges support by the Swiss National Science Foundation, grant PP00P2\_163824.
GI is supported by the Royal Society Newton International Fellowship (NF170902).

%%-------------------------------------------------------------------
\bibliographystyle{aa}
\bibliography{biblio.bib}
%%-------------------------------------------------------------------

\appendix
\section{The Milky Way gas distribution compared with the literature}\label{ap:gas_distrib}
In the following, we compare the radial profiles of the surface density, the volume density, and the scale height adopted in this work with those available in the 
literature, for the HI, the H$_2$, and the total gas. 

\subsection{Gas surface densities}\label{ap:Sigmagas_compare}
\begin{figure*}
\includegraphics[width=2.\columnwidth]{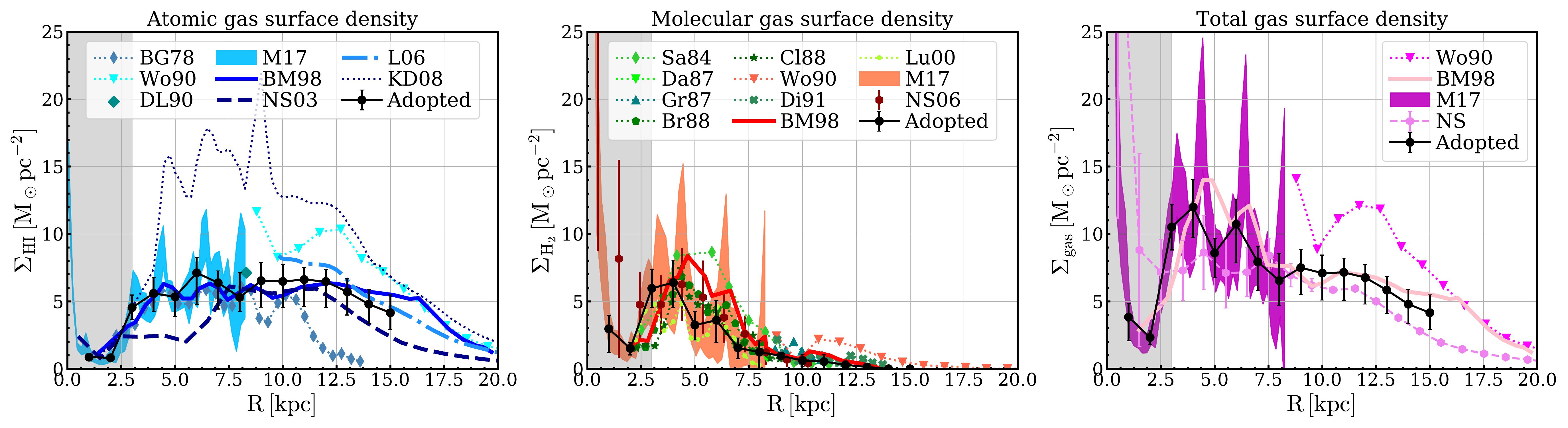}
\caption{Surface density radial profiles of the atomic gas (left), the molecular gas (center), and the total gas (right) from the literature compared to the one adopted in this 
work (black points). 
The factor 1.36 for the helium correction is included. 
In the left panel: BG78=\cite{1978BurtonGordon}, Wo90=\cite{1990Wouterloot}, DL90=\cite{1990DickeyLockman}, BM98=\cite{1998BinneyMerrifield}, NS03=\cite{2003NakanishiSofue}, 
Le06=\cite{2006Levine}, KD08=\cite{2008KalberlaDedes}, and M17=\cite{2017Marasco}.
In the central panel: Sa84=\cite{1984Sanders}, Da87=\cite{1987Dame}, Gr87=\cite{1987Grabelsky}, Br88=\cite{1988Bronfman}, Cl88=\cite{1988Clemens}, Di91=\cite{1991Digel}, 
Lu00=\cite{2006Luna}, NS06=\cite{2006NakanishiSofue}.}
\label{fig:Sigma_gas}
\end{figure*}

Fig.~\ref{fig:Sigma_gas} shows the radial profiles of the surface density of the atomic gas (left), the molecular gas (center) and the total gas (right) in the MW (all profiles 
are rescaled to the same value for $R_\odot$). 
The black points represent the profiles adopted in this work, that were calculated using profiles from the literature smoothed to 1 kpc resolution. 
For example, at 8 kpc, we used the measurements of the surface density from 7.5 kpc to 8.5 kpc and fitted a Gaussian distribution to the data (accounting for the errors). 
The resulting best-fit Gaussian is centered on the final value for the density, i.e. the black point at 8 kpc, and its standard deviation is a first estimate of the error. 
We adopted the same procedure for both HI and H$_2$ profiles. 
These fiducial profiles are compared to those in the literature, whose reference papers are indicated by the initials in the legend of Fig.~\ref{fig:Sigma_gas}.  

Within the Solar circle, we used $\Sigma_\mathrm{HI}$ and $\Sigma_\mathrm{H_2}$ from \cite{2017Marasco}. 
The upper errors on $\Sigma_\mathrm{HI}$ were estimated from the standard deviation of the Gaussian, while the lower errors were found using $\Sigma_\mathrm{HI}$ derived 
in optically thin regime (see \citealt{2017Marasco} for details). 
This sort of correction was done for consistency with B19, in which we used $\Sigma_\mathrm{HI}$ of external galaxies derived under the assumption of 100\% optical thin HI. 
Concering the molecular gas, we followed \cite{2017Marasco} approach and associated an error of 30\% to $\Sigma_\mathrm{H_2}$ based on the uncertainty on 
the CO-H$_2$ conversion factor estimated by \cite{2013Bolatto}.

Beyond the Solar circle, we adopted $\Sigma_\mathrm{HI}$ profiles from \cite{1998BinneyMerrifield}, \cite{2003NakanishiSofue}, and \cite{2006Levine}. 
These authors assumed an opaque regime for the HI, but did not explore the optically thin case, thus we used the Gaussian standard deviation for the error. 
Similarly, $\Sigma_\mathrm{H_2}(R)$ was derived using the estimates by \cite{2006NakanishiSofue} and \cite{1998BinneyMerrifield} and we associated a 30\% uncertainty based 
on \cite{2013Bolatto}. 

The most uncertain profile is probably $\Sigma_\mathrm{HI}$ (see Fig.~\ref{fig:Sigma_gas}), while the different estimates of $\Sigma_\mathrm{H_2}$ seem approximately compatible. 
The discrepancies between the HI profiles could be due to the differences in the assumed rotation curve, that has an important effect in the kinematic distance method 
\citep[see][]{1974Burton,2017Marasco}. 

\subsection{Gas scale height}\label{ap:hgas_compare}
\begin{figure*}
\includegraphics[width=2.\columnwidth]{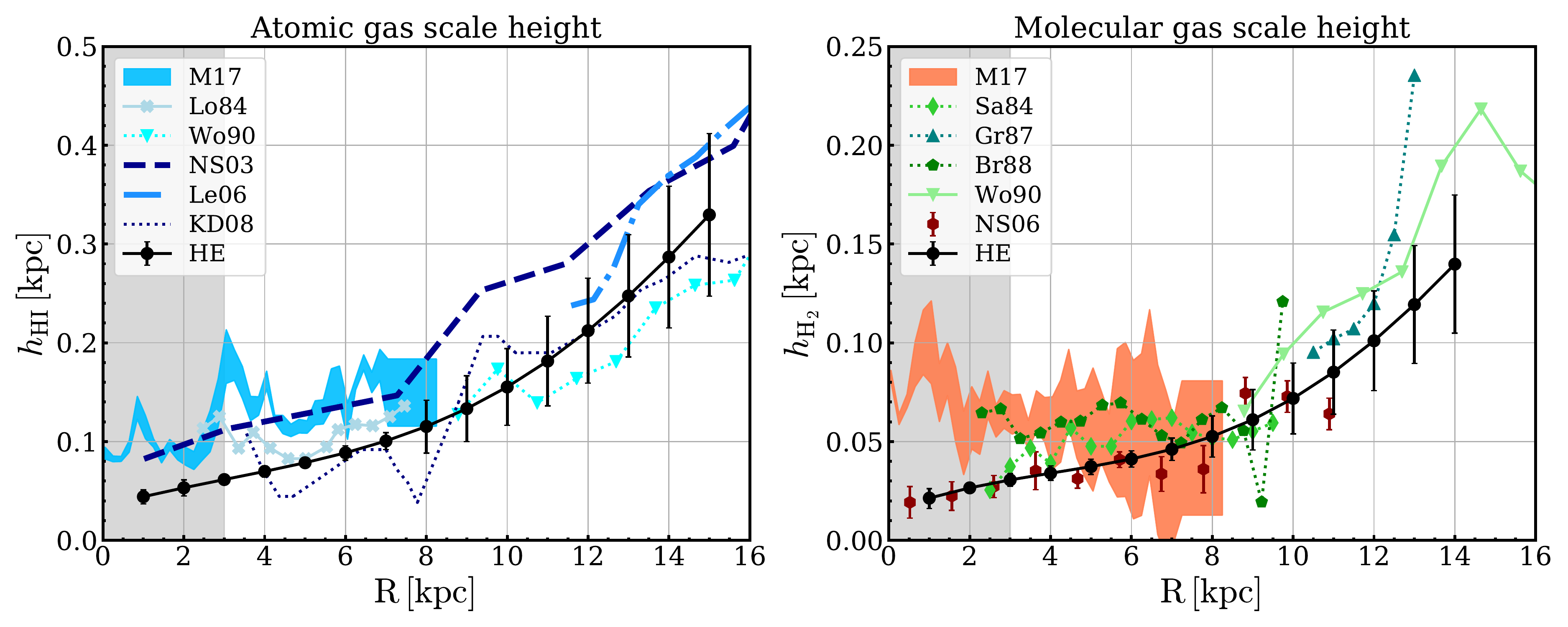}
\caption{Scale height radial profiles of the atomic and molecular gas from the literature (Lo84=\citealt{1984Lockman}; see caption of Fig.~\ref{fig:Sigma_gas} for the other initials 
in the legend) compared to those calculated assuming the hydrostatic equilibrium (black points).} 
\label{fig:h_gas}
\end{figure*}

There have been several attempts to infer the gas scale height in our Galaxy directly from the data, and it is interesting to compare these determinations with those we calculated  
with the hydrostatic equilibrium (see Sect.~\ref{sec:h_gas_hydro}). 
These latter are represented by the black points in Fig.~\ref{fig:h_gas}, where the left and the right panel show respectively the HI and the H$_2$ scale heights. 

Despite the very different methods, all the profiles show a flaring. 
The scale height of HI in hydrostatic equilibrium is compatible with the results of \cite{1984Lockman}, \cite{1990Wouterloot}, \cite{2006Levine} and \cite{2008KalberlaDedes}. 
This latter in particular assumed the hydrostatic equilibrium a priori and modelled the ISM as a two-phases fluid, in which the warm neutral medium (WNM) and the cold neutral 
medium (CNM) have different scale heights. 
They found that the final model is close to a single-component medium with constant velocity dispersion at 8.3 \kms, that is similar to the values we adopted for the HI. 
This can explain the agreement between the scale heights, despite the significant differences between the mass models \citep[see][]{2007Kalberla}. 

However, our scale height differs from those derived by \cite{2003NakanishiSofue}\footnote{\cite{2003NakanishiSofue} assumed a $\sech^2$ profile for the HI vertical distribution, 
so we rescaled their HWHM by a factor 0.8 to obtain the equivalent quantity for a Gaussian profile.} and \cite{2017Marasco} by $\sim 50$\% (see left panel 
in Fig.~\ref{fig:h_gas}). 
\cite{2003NakanishiSofue} adopted the kinematic distance method to measure the HI distribution and removed the emission of high-altitude and diffuse HI, so a direct 
comparison is not straighforward. 
The discrepancy with \cite{2017Marasco} scale height deserves further discussions, as we adopted their velocity dispersion profile to calculate $h_\mathrm{HI}$, 
hence we expected the scale heights to be compatible within the uncertainties. 
There are some possible explanations for this difference. 
\cite{2017Marasco} found that HI in the midplane could be best described by a two-component model, where the 80\%-85\% of the atomic gas has low velocity dispersion ($\sim 8$ \kms), 
while the remaining has a much higher velocity dispersion (15-20 \kms). 
Likely, this second component has a larger scale height than the first one, in which case it could dominate the HI emission above the midplane. 
The vertical distribution of the gas was instead modelled using a single component, thus it is possible that the resulting $h_\mathrm{HI}$ is closer to the scale height of 
the high-$\sigma_\mathrm{HI}$ component than to that of the low-$\sigma_\mathrm{HI}$ one. 
We adopted the measured velocity dispersion to calculate the scale height with hydrostatic equilibrium, but this approach implicitly relies on the assumption 
that the velocity dispersion is isotropic and constant along $z$. 
However, this latter property may not be true if the high-$\sigma_\mathrm{HI}$ component is more abundant than the low-$\sigma_\mathrm{HI}$ one at high 
latitudes above the midplane. 
Alternatively, we could speculate that some anisotropic force contributes to balance the gravitational pull towards the midplane, like magnetic tension or cosmic rays 
for instance. 
In particular, recent magneto-hydrodynamical simulations of stratified gas of galaxies \citep[e.g.][]{2016Simpson,2017Pfrommer} showed that the anisotropic diffusion of cosmic 
rays can contribute to the vertical gradient of the gas pressure, but investigating such scenarios is beyond the aim of this paper. 

Fig.~\ref{fig:h_gas} shows that the molecular gas scale height based on the hydrostatic equilibrium resembles the $h_\mathrm{H_2}$ by \cite{2006NakanishiSofue} and 
\cite{2017Marasco} within the uncertainties. 
Also $h_\mathrm{H_2}$ from \cite{1984Sanders}, \cite{1987Grabelsky}, and \cite{1990Wouterloot} are in approximate agreement with our scale height, 
while \cite{1988Bronfman} scale height shows some discrepancies for 3 kpc $\lesssim R \lesssim$ 8 kpc. 

\subsection{Gas volume densities}\label{ap:rhogas_compare}
\begin{figure*}
\includegraphics[width=2.\columnwidth]{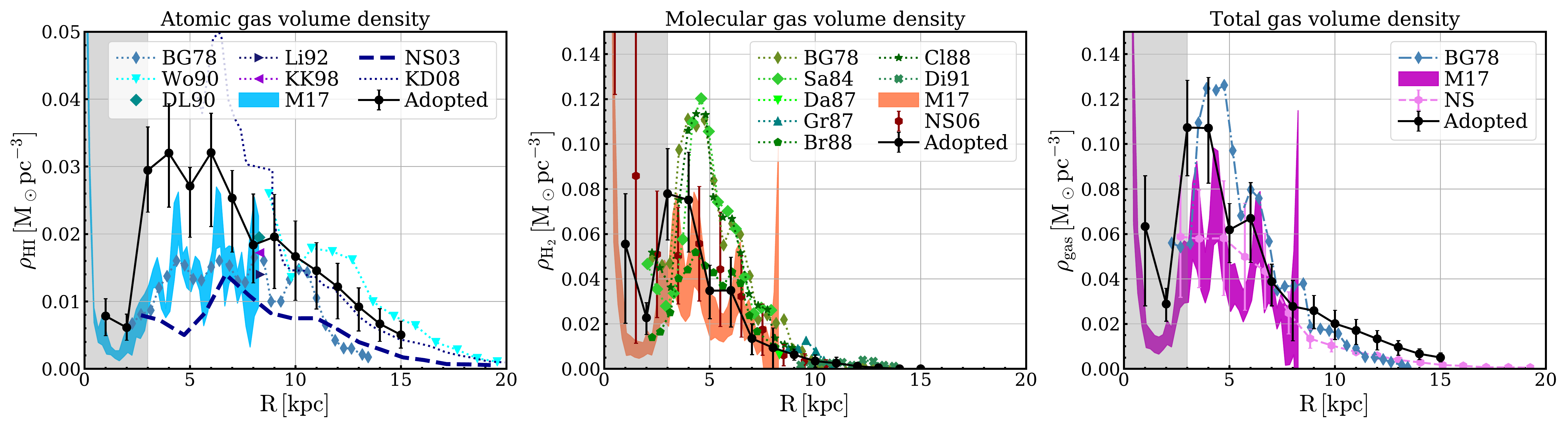}
\caption{Volume density radial profiles of the atomic gas (left), the molecular gas (center), and the total gas (right) from the literature (Li92=\citealt{1992Liszt}, 
KK98=\citealt{1998KalberlaKerp}, NS=HI from \citealt{2003NakanishiSofue} and H$_2$ from \citealt{2006NakanishiSofue}; see caption of Fig.~\ref{fig:Sigma_gas} for the other 
initials in the legend) compared to the one adopted in this work (black points). 
The factor 1.36 for the helium correction is included.}
\label{fig:rho_gas}
\end{figure*}

In Fig.~\ref{fig:rho_gas}, the black points represent the MW volume density profiles of atomic gas (left panel), molecular gas (central panel), and total gas (right panel)
calculated with the scale height of hydrostatic equilibrium. 
The other points show instead the measurements available in the literature. 

Within the Solar radius, our $\rho_\mathrm{HI}$ is systematically higher than the other estimates, as we expected from the discrepancy found between the scale heights. 
However, this difference seems to be partially accounted for by the errorbars, that were calculated from the uncertainties on $h_\mathrm{HI}$ and $\Sigma_\mathrm{HI}$. 
These latter in particular include the difference between the profiles in the literature (beyond $R_\odot$) and the uncertainty on the optical regime of atomic gas 
(within $R_\odot$), as reported in Sect.~\ref{ap:Sigmagas_compare}. 
Concering the molecular gas, our profile for $\rho_\mathrm{H_2}(R)$ is in agreement with the observations and, given that the gas is mainly molecular within the Solar circle, 
also our profile for the total gas is approximately compatible with all the other determinations. 

\section{The Milky Way SFR distribution using different tracers}\label{ap:sfr_tracers}
In the following, we compare the surface density and the scale height of the SFR adopted in this work with other profiles available in the literature. 

\subsection{SFR surface density}\label{ap:sfr_tracers_Sigma}
We adopted the SFR surface density profile derived by \cite{2015Green} using SNRs to trace the distribution of recent star formation. 
In the left panel of Fig.~\ref{fig:Sigma_h_SFR}, this profile is compared to others in the literature, all normalised to have a total SFR of $1.9$ M$_\odot$yr$^{-1}$. 
The discrepancy with \cite{1998CaseBhattacharya}, who also studied the SNR distribution, was expected (see \citealt{2015Green}). 
The adopted $\Sigma_\mathrm{SFR}(R)$ is compatible with the profile derived using pulsars \citep{1985Lyne,2004Yusifov} and the far-infrared emission \citep{2006Misiriotis}, 
except for the inner 3 kpc that are not included in our study. 
\cite{2015Green}'s profile is also compatible with the HII regions radial distribution \citep{2004Paladini}, except for $R\sim 5-6$ kpc (see Appendix~\ref{ap:vsflaw_others} 
for further discussion). 

\subsection{SFR scale height}\label{ap:sfr_tracers_h}
We studied the radial profile of CCs scale height using the residual $z$-coordinate ($\Delta z$) provided by \cite{2019Chen} (see their Fig. 4), for which the signature of 
the Galactic warp was already modelled and filtred out. 
We built radial bins of $\Delta R =1$ kpc from $R=5$ kpc to $R=19$ kpc cointaining enough stars (from 8 to 200 stars) to obtain a reasonable sampling of their vertical 
distribution in each bin using the Freedman--Diaconis estimator \citep{Freedman81onthe} implemented in the \texttt{scipy} Python package \citep{scipy}. 
By analogy with a Gaussian distribution, we calculated the scale height at each radius, i.e. the width of the vertical distribution in each bin, as 
\begin{equation}
 h_\mathrm{Cep} = \frac{p_\mathrm{84}-p_\mathrm{16}}{2} \, ,
\end{equation}
with $p_\mathrm{16}$ and $p_\mathrm{84}$ being the 16th and the 84th percentiles of the distribution in the bin. 
The uncertainty on this estimate is the sum in quadrature of two contributions:
\begin{equation}
 \Delta h_\mathrm{Cep} = \left[ \left(p_\mathrm{50} - 
 \frac{ p_\mathrm{84} + p_\mathrm{16} }{2} \right)^2 + 
 \left( \frac{h_\mathrm{Cep}}{\sqrt{N}} \right) ^2 \right]^{\frac{1}{2}} \, ,
\end{equation}
where $p_\mathrm{50}$ and $N$ are respectively the median of the distribution, i.e. the midplane, and the number of CCs in each bin. 
The first term comes from the asymmetry with respect to the midplane (with the warp contribution already subtracted), while the second term accounts for the statistical error. 

The resulting scale height is shown in Fig.~\ref{fig:h} and it is compatible with that reported by \cite{2019Chen}, despite the different definitions adopted. 
As a further test, we took the catalogue of CCs by \cite{2019Skowron}, who estimated the age of each star, and selected the youngest population (20 Myr$<$ age $\leq$ 90 Myr). 
For this latter, we calculated the scale height and found that it is compatible with that shown in Fig.~\ref{fig:h} for \cite{2019Chen}'s sample. 
Moreover, the scale of young CCs is the same as that estimated by \cite{2019Skowron} for the sample including all ages, which indicates that CCs scale height does not depend 
significantly on age.

\begin{figure*}
\includegraphics[width=2.\columnwidth]{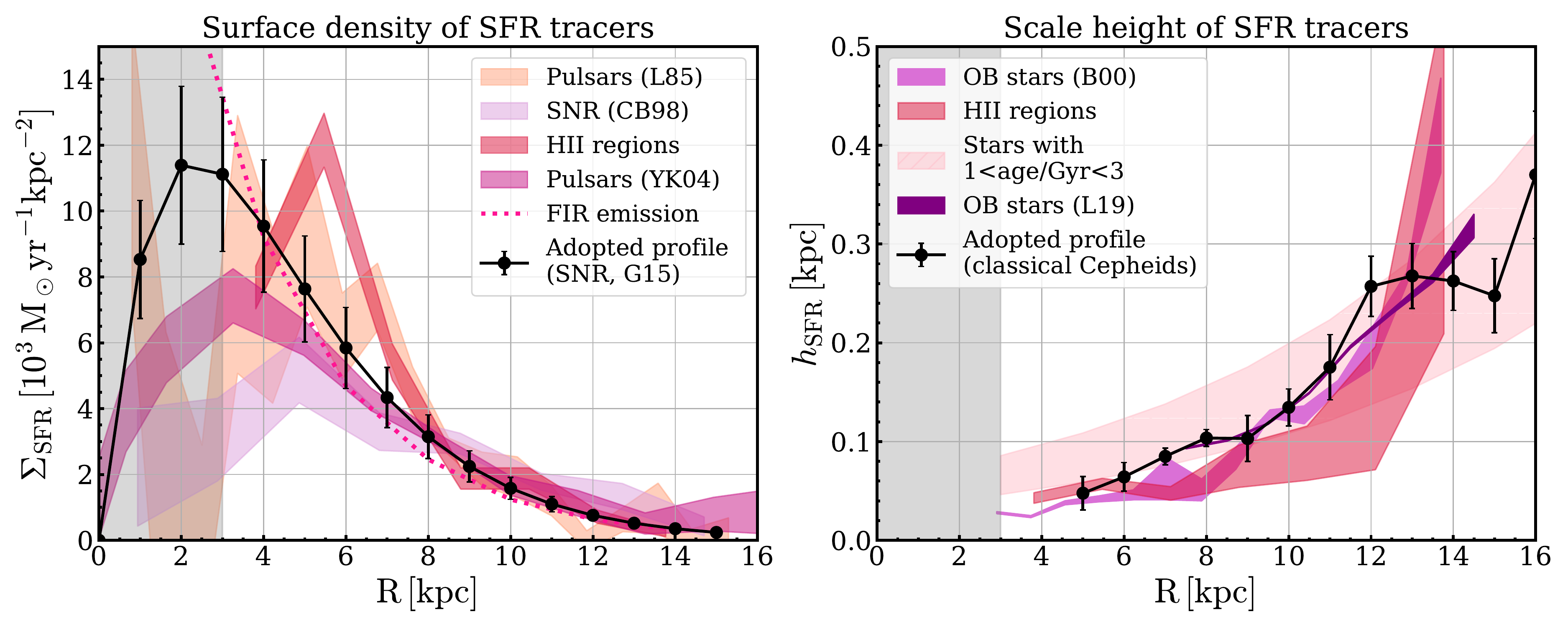}
\caption{Surface density (left panel) and scale height (right panel) of the SFR distribution using different tracers. 
The black points show the profile adopted in this work, while the others are different estimates from the literature (see text). 
The reference papers are indicated by the following abbreviations: L85=\cite{1985Lyne}, CB98=\cite{1998CaseBhattacharya}, YK04=\cite{2004Yusifov}, G15=\cite{2015Green}, 
B00=\cite{2000Bronfman}, and L19=\cite{2019Li}.} 
\label{fig:Sigma_h_SFR}
\end{figure*}

The right panel in Fig.~\ref{fig:Sigma_h_SFR} shows the comparison between the scale height of CCs (black points) with the scale height of other tracers of 
star formation: OB stars \citep{2000Bronfman,2019Li}, stellar populations with 1 Gyr < age < 3 Gyr \citep{2017Mackereth}, and HII regions \citep[][Table 4]{2004Paladini}. 
\cite{2017Mackereth} and \cite{2019Li} assumed an exponential function to model the vertical profile of young stars, thus we rescaled their profiles by a factor 
$1/\sqrt{2}$ in order to be comparable with our definition of the scale height, which is equivalent to the normalised second moment of the distribution. 
The presence of the flare is beyond any doubt in all the cases and, despite the different methods adopted, the profiles are generally compatible. 

\section{VSF law with alternative determinations from the literature}\label{ap:vsflaw_others}
\begin{figure*}
\includegraphics[width=2.\columnwidth]{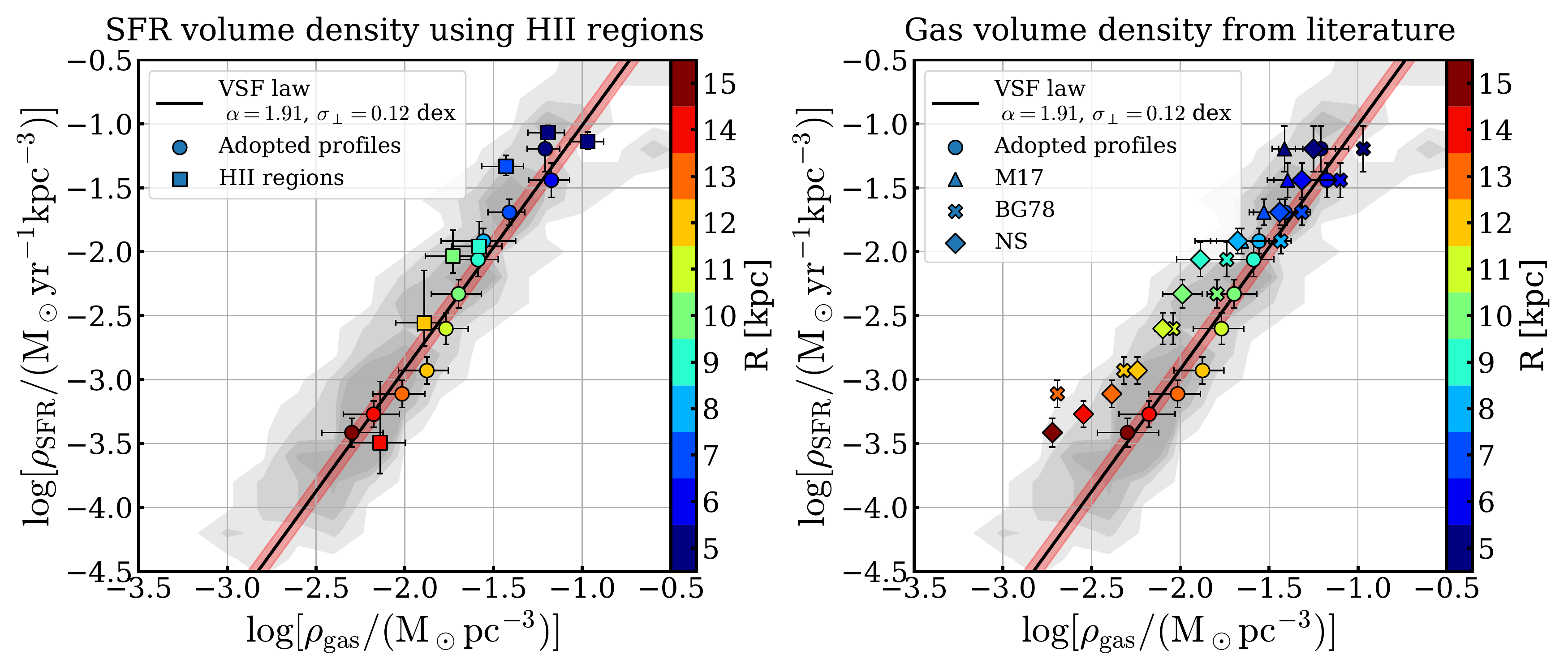}
\caption{Left panel: comparison between $\rho_\mathrm{SFR}$ derived with the surface density of SNRs and the scale height of CCs 
(circles, same as Fig.~\ref{fig:vsf}) and $\rho_\mathrm{SFR}$ calculated with the surface density and the scale height of HII regions (squares). 
In both cases, $\rho_\mathrm{gas}$ is calculated with hydrostatic equilibrium (see Sect.~\ref{sec:gas_dist_kin}). 
Right panel: comparison between $\rho_\mathrm{gas}$ derived with the hydrostatic equilibrium (circles, same as Fig.~\ref{fig:vsf}) and $\rho_\mathrm{gas}$ from different works 
in the literature (M17=\citealt{2017Marasco}, BG78=\citealt{1978BurtonGordon}, NS=\citealt{2003NakanishiSofue,2006NakanishiSofue}), all with $\rho_\mathrm{SFR}$ used in this work 
(see Sect.~\ref{sec:sfr_distribution}). 
In both panels, the symbols are colour-coded according to the Galactocentric radius and the solid line and the red band show the VSF law and its intrinsic scatter derived 
for nearby disc galaxies, whose volume densities are represented by the grey contours cointaining, from the lightest to the darkest, the 95\%, 75\%, 50\%, and 25\% 
of the points.}
\label{fig:vsf_obs}
\end{figure*}

Here we build the VSF law for the MW using other estimates of the gas and the SFR volume densities. 
The aim of this exercise is to demonstrate that the MW is compatible with the VSF law no matter which measurement we choose or whether the hydrostatic equilibrium is assumed. 
In both panels of Fig.~\ref{fig:vsf_obs}, the circles show the profile adopted in this work (see Sect.~\ref{sec:gas_dist_kin} and Sect.~\ref{sec:sfr_distribution}) and are the 
same as in Fig.~\ref{fig:vsf}. 
In the left panel, the squares indicate $\rho_\mathrm{SFR}$ calculated using the surface density and the scale height radial profiles of HII regions provided by \cite{2004Paladini}. 
The two determinations are well compatible except for the points at $R\sim 7$ kpc, where there is a gap between the scale height of CCs and that of HII regions, 
and also the surface densities are marginally different (see Fig.~\ref{fig:Sigma_h_SFR}). 
The agreement between the two determinations is remarkable and significantly consolidates our results. 

In the right panel, we show different $\rho_\mathrm{gas}$ taken from the exhaustive collection in \cite{2016KramerRandall}, who provide the volume densities of the atomic and the 
molecular gas by \cite{1978BurtonGordon} and \cite{2003NakanishiSofue,2006NakanishiSofue} (the uncertainties are unfortunately not available), and from \cite{2017Marasco}. 
All the points in Fig.~\ref{fig:vsf_obs} are colour-coded according to the Galactocentric radius and the spread in the points with the same colour, i.e. at the same $R$, 
can be interpreted as the uncertainty on the volume density measurements at that radius. 
We also show the volume densities of the sample of galaxies in B19 (grey contours) and the VSF law with slope $\alpha=1.91 \pm 0.03$. 
The MW points are generally compatible with those of nearby galaxies and the VSF law, although the large scatter makes the comparison itself uncertain. 
The MW points derived without using the assumption of hydrostatic equilibrium seem to suggest a shallower slope ($\sim 1.5$) for the VSF law, but again the large uncertainties do 
not allow to draw robust conclusions. 

\end{document}